\documentclass{desyproc}
\usepackage{subfigure}
\usepackage{hyperref}
\begin{document}
\title{Nucleon Transverse Structure at COMPASS}

\author{{\slshape Nour Makke$^{1}$}\\[1ex]
$^1$Trieste University and INFN section of Trieste, Via A.~Valerio 2, 34127 Trieste, Italy} 

\contribID{140}

\confID{8648}  
\desyproc{DESY-PROC-2014-04}
\acronym{PANIC14} 
\doi  

\maketitle

\begin{abstract}
COMPASS is a fixed target experiment at CERN. Part of its physics programme is dedicated to study the transverse spin and the transverse momentum structure of the nucleon using SIDIS. For these measurements, data have been collected using transversely polarised proton and deuteron targets. A selection of recent measurements of azimuthal asymmetries using data collected with transversely polarised protons is presented.
\end{abstract}

\section{Introduction}
\label{sec:intro}

The description of the partonic structure of the nucleon remains one of the main challenges in hadron physics. In the last few decades, a considerable theoretical and experimental progress has been accomplished and the relevance of the quark transverse spin and transverse momentum to study its structure has been assessed. The nucleon constituents are not only collinear moving objects but they may also have a momentum component transverse to the nucleon direction of motion.
In the present theoretical framework, eight transverse momentum dependent parton distribution functions (TMD PDFs) are required for each quark flavour at leading twist, describing all possible correlations between the transverse momentum and spin of the quarks, and the spin of the nucleon. The most famous and studied one is the Sivers PDF. Integrating over the quark intrinsic transverse momentum, five among these functions vanish and three survive giving the well known number ($q(x,Q^2)$), helicity ($\Delta q(x,Q^2)$) and transversity ($\Delta_{\perp} q(x,Q^2)$) distribution functions. Experimentally, the latter is the least known one. Beside these, many other twist-2 distributions can be introduced, correlating the spin and the transverse momentum.  

Many processes are being, and will be, studied to access the TMD PDFs, namely transversely polarised hard proton-proton collisions, Drell-Yan processes and semi-inclusive deep inelastic scattering (SIDIS). Although they are complementary, the last channel is nowadays the major source of information. Its main advantage is that TMD effects generate different azimuthal modulations in its cross section, which can be separately explored and extracted from the same experimental data set. The modulations depend on two angles, $\phi_S$ and $\phi_h$ which define the azimuthal angle of the initial nucleon spin and the produced hadron momentum respectively. These angles are defined in the so called gamma nucleon system in which the direction of the virtual photon is the $z$ axis and  the $xz$ plane is defined by the lepton scattering plane. The modulation amplitudes are different structure functions, proportional to the convolution of the TMD PDFs and fragmentation functions (FFs).

The transversity distributions $\Delta_{\perp} q(x)$ can not be measured in inclusive DIS due to their chirally odd nature. They can instead be extracted from measurements of single-spin azimuthal asymmetries in cross-sections for SIDIS of leptons on transversely polarised nucleons, in which a hadron is also detected. The measurable asymmetry, the Collins asymmetry, is due to a combined effect of $\Delta_{\perp}q$ and the chiral-odd Collins TMD-FF $\Delta_{T}^{0}D_{q}^{h}$, which describes the spin-dependent part of the hadronization of a transversely polarised quark into a hadron with transverse momentum $p_T$. At leading order, the Collins mechanism leads to a modulation in the azimuthal distribution of the produced hadrons given by:

\begin{equation}
N(\phi_C) = \alpha(\phi_C) \cdot N_0 (1 + A_{Coll} \cdot P_{T} \cdot f \cdot D_{NN} \sin\phi_C)
\end{equation}

where $\alpha$ contains the apparatus efficiency and acceptance, $P_T$ is the target polarisation, $D_{NN}$ is the spin transfer coefficient and $f$ is the fraction of polarisable nuclei in the target, $\phi_C = \phi_h - \phi_{S'} = \phi_{h} + \phi_{S} - \pi$ is the Collins angle, with $\phi_h$ the hadron azimuthal angle, $\phi_{S'}$ the final azimuthal angle of the quark spin and $\phi_S$ the azimuthal angle of the nucleon spin in the gamma-nucleon system. 

\section{The COMPASS experiment}
\label{sec:spectrometer}

COMPASS~\cite{0} (COmmon Muon and Proton Apparatus for Structure and Spectroscopy) is a fixed target experiment located at the CERN SPS taking data since 2002. Semi-inclusive DIS data have been collected using a 160 GeV longitudinally polarised muon beam and longitudinally or transversely polarised proton (NH$_3$) and deuteron ($^6$LiD) targets. The spectrometer comprises a variety of different tracking detectors, and allows to detect charged tracks in a broad momentum and angular range. Calorimeters, muon filters and a gas radiator RICH detector are available for particle identification.

\section{Data Analysis and Results}
\label{sec:results}

Collins and Sivers asymmetries have been extracted as a function of $x$, $z$ and $p_T$ for positive and negative hadrons, pions and kaons, using lepton scattering off transversely polarised deuterons (2002-04) and protons (2007,2010). Using a deuteron target, the resulting Collins and Sivers asymmetries turned out to be compatible with zero~\cite{1},\cite{2}, an observation which has been interpreted as a cancellation between the $u$ and $d$ quark contributions in the isoscalar target. Using a proton target, a first measurement was performed separately versus $x$, $z$ and $p_T$ for unidentified hadrons, pions and kaons ~\cite{3},\cite{4},\cite{5}. The Collins asymmetry is small, compatible with zero, except for $x \ge 0.05$ where a significant signal (up to 10 \%) appears with opposite sign for positive and negative hadrons, pions and kaons. The results for the Sivers asymmetry are compatible with zero for negative hadrons and exhibit small positive values (up to 3\%) for positive hadrons both at small and at large $x$. Compared with HERMES results measured at smaller $Q^2$, the results were found to be slightly smaller. A further investigation showed that the signal is concentrated at small $W$ while at larger $W$, it tends to zero. Thus COMPASS data highlights a possible $W$ dependence of the Sivers asymmetry for positive hadrons. The other six asymmetries were extracted from deuteron and proton data and were found to be compatible with zero.

Recently, further investigation of the previous observations has been performed by studying the $x$, $z$, $p_T$ and $W$ dependencies in different $Q^2$ regions: $Q^2 \in$ [1,4], [4,6.25], [6.25,16] and $Q^2 \ge 16$ (GeV/c)$^2$, using the data set collected in 2010 on a transversely polarised proton. 

\begin{figure}[htdp]
\centerline{
\includegraphics[width=.95\textwidth]{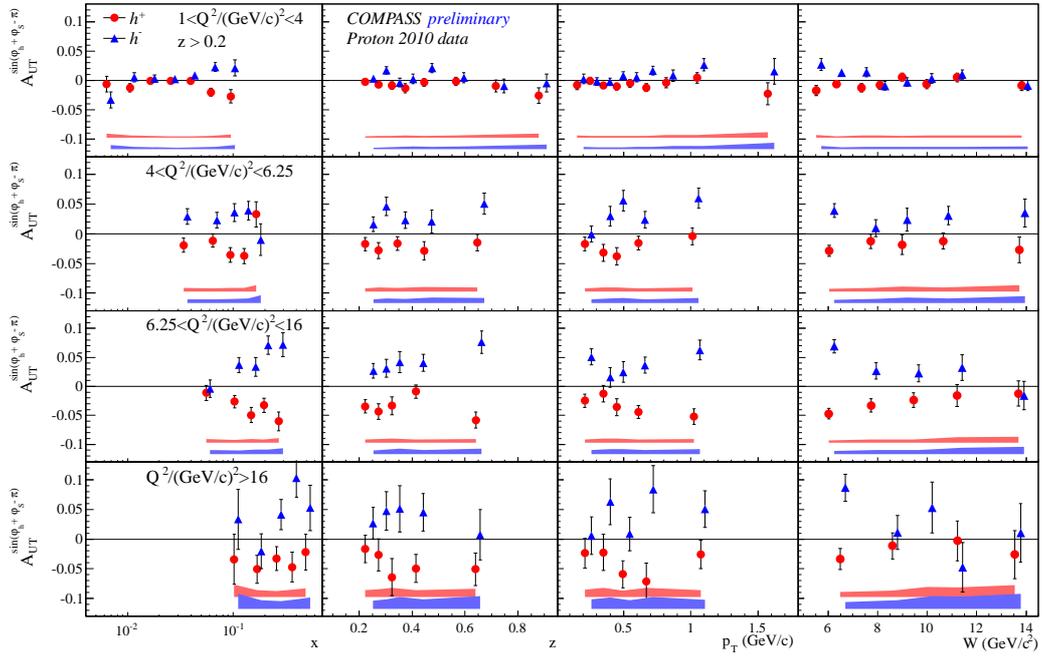}}
\caption{Collins asymmetry vs. $x$, $z$, $p_T$, $W$ in four $Q^2$ ranges.}\label{Fig:Col}
\end{figure}
\begin{figure}[htdp]
\centerline{
\includegraphics[width=.95\textwidth]{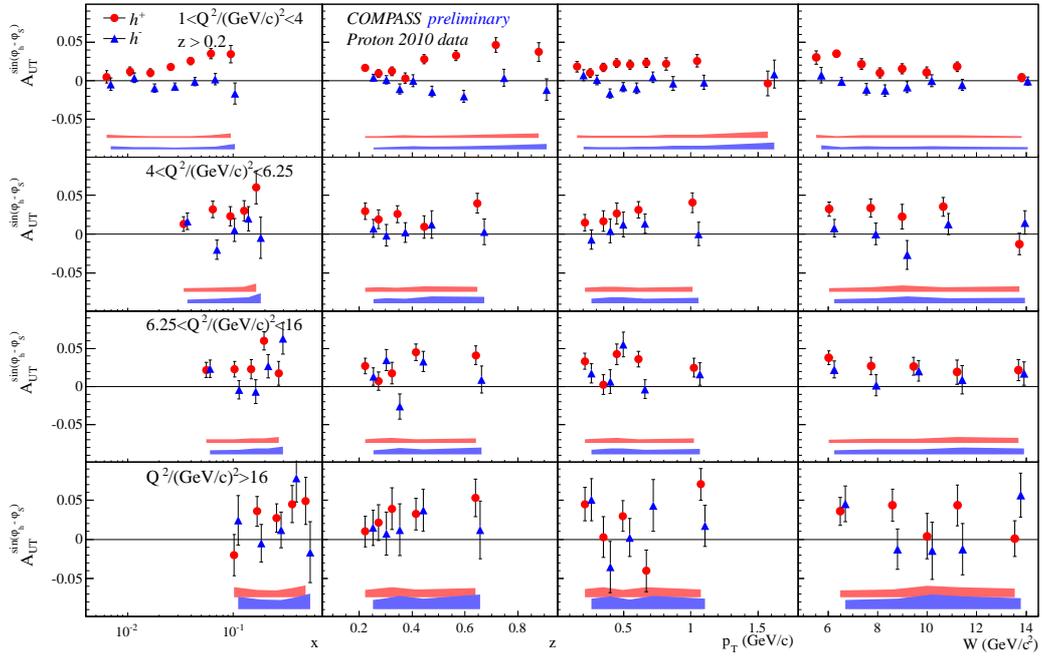}}
\caption{Sivers asymmetry vs. $x$, $z$, $p_T$, $W$ in four $Q^2$ ranges.}\label{Fig:Siv}
\end{figure}
\begin{figure}[htdp]
\centerline{
\subfigure{\includegraphics[width=0.28\textwidth]{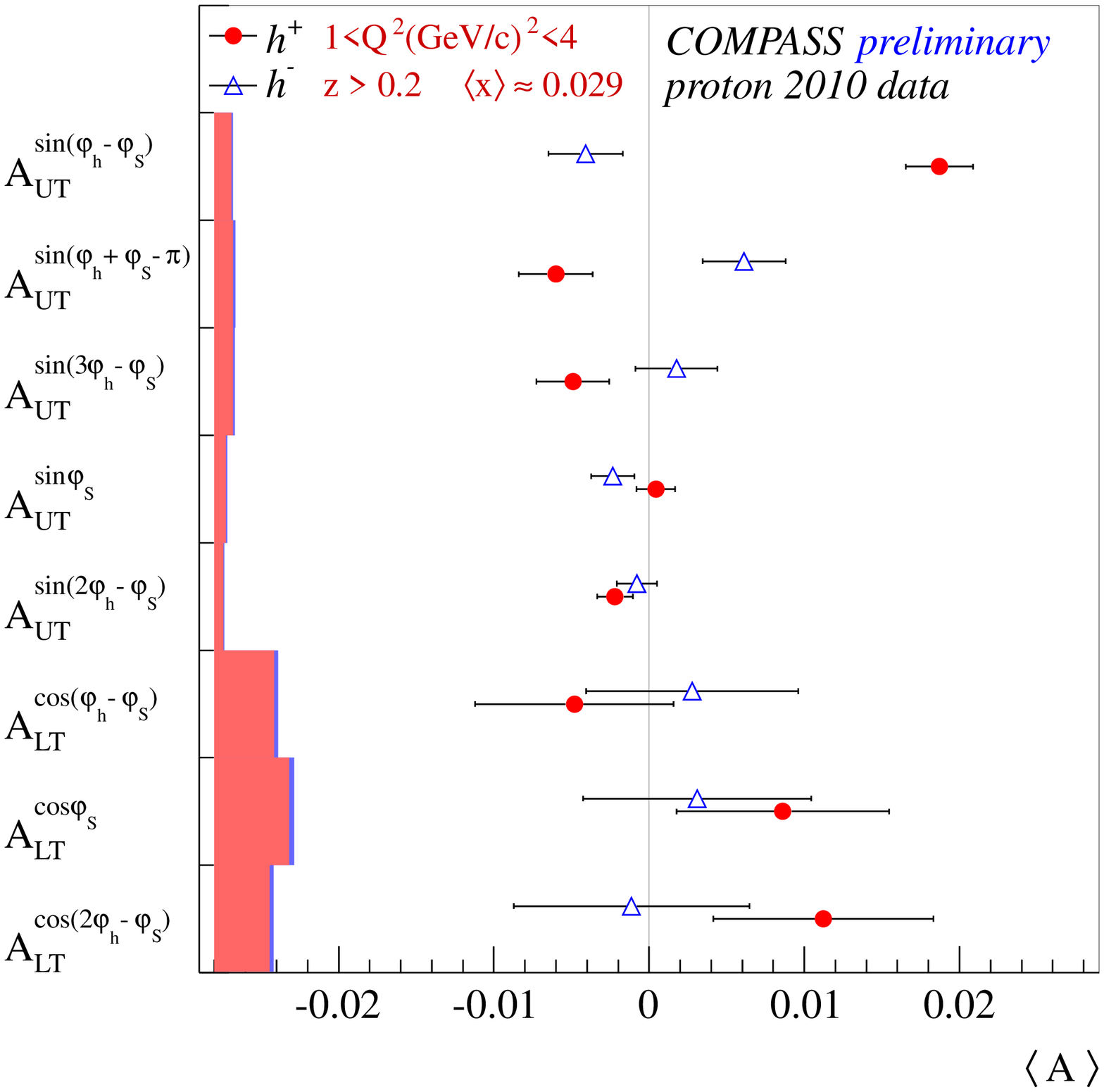}}\subfigure{\includegraphics[width=0.28\textwidth]{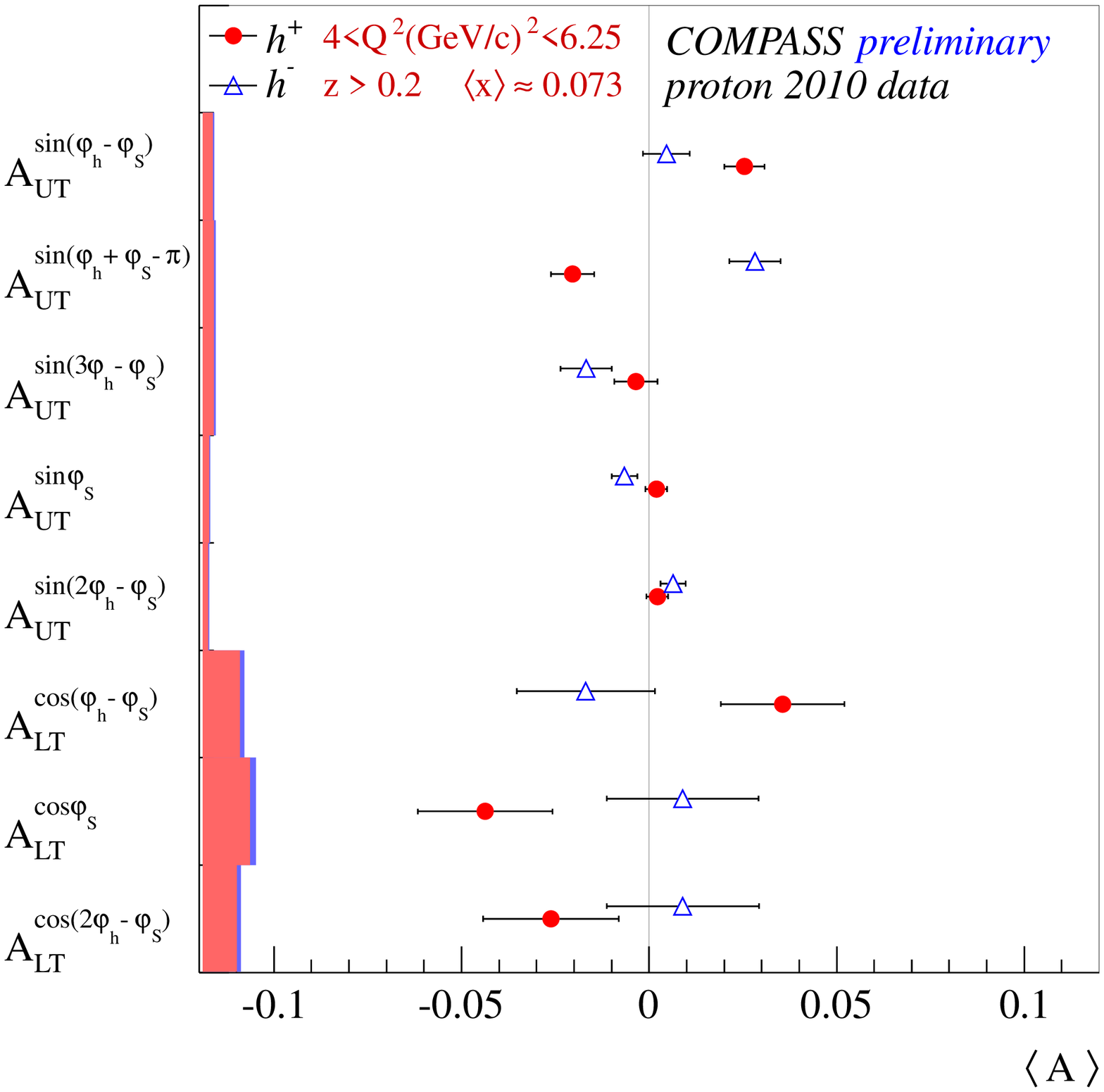}}}
\end{figure}
\begin{figure}[htdp]
\vspace{-0.6cm}
\centerline{
\subfigure{\includegraphics[width=0.28\textwidth]{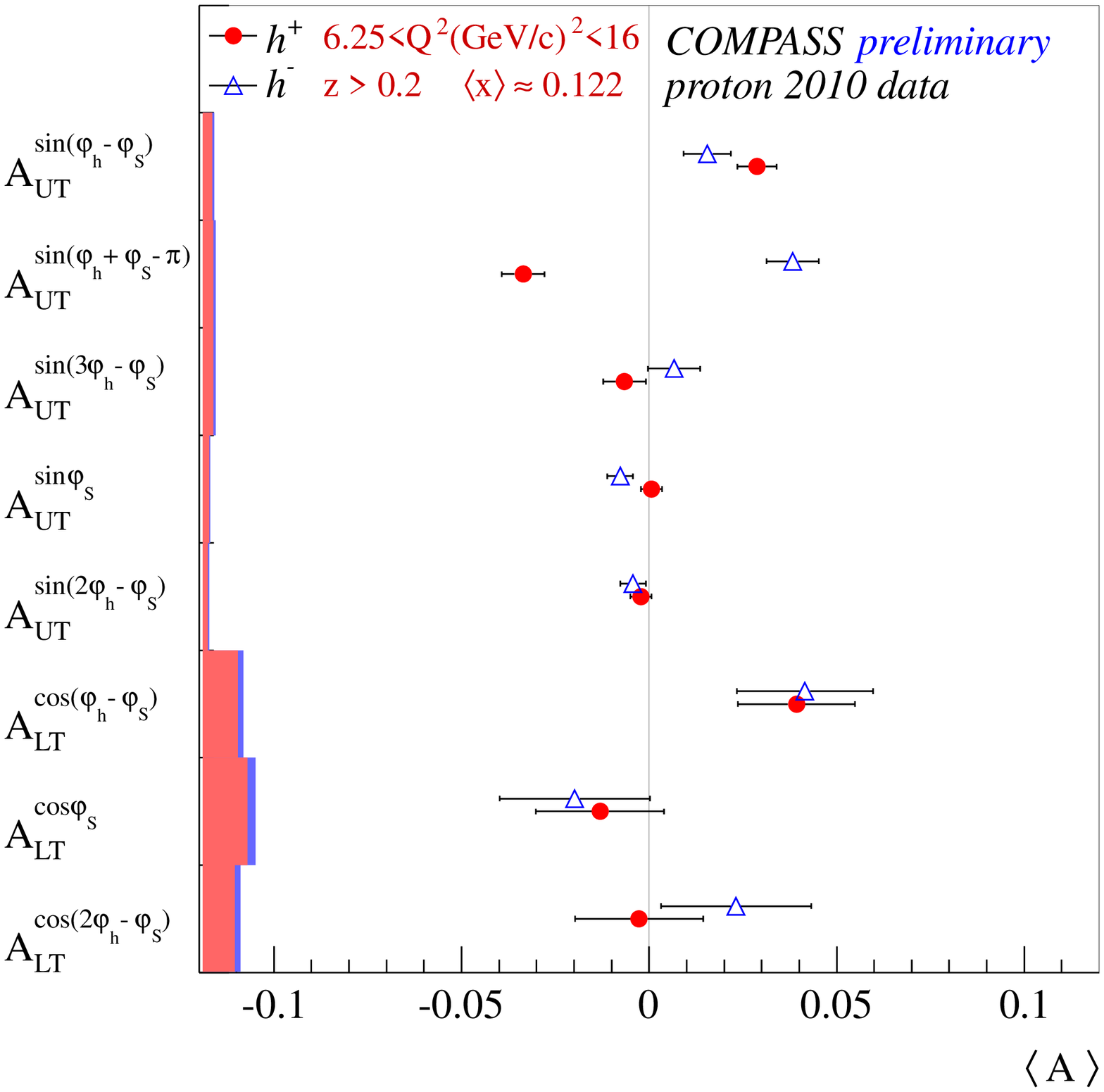}}
\subfigure{\includegraphics[width=0.28\textwidth]{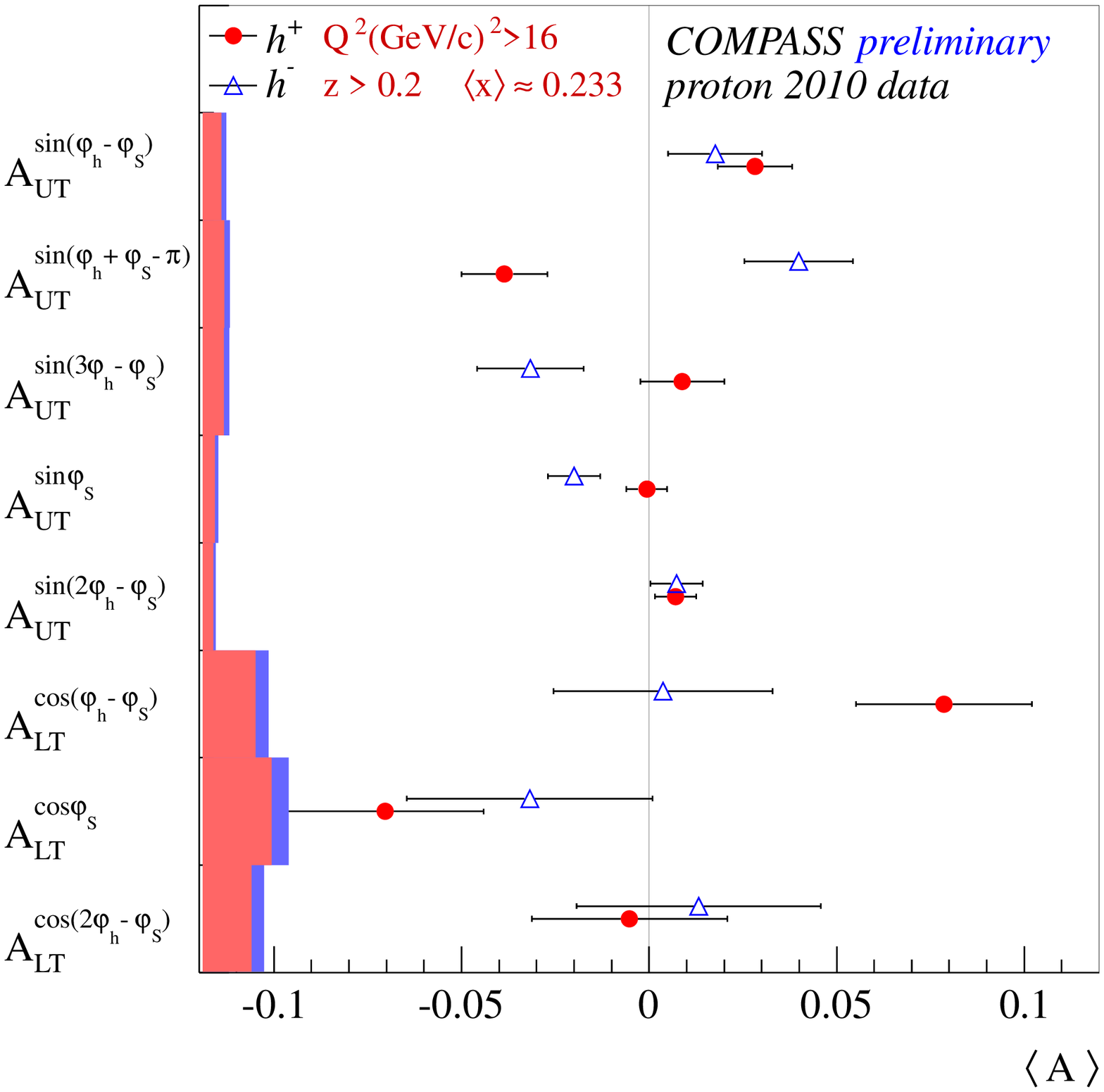}}
}
\caption{Mean of the eight asymmetries in four $Q^2$ regions.}\label{Fig:All}
\end{figure}

These new results for Collins and Sivers asymmetries are shown in Fig.~\ref{Fig:Col}-\ref{Fig:Siv} versus $x$, $z$, $p_T$ and $W$ and the mean of the eight asymmetries are shown in Fig.~\ref{Fig:All}. Error bars show only statistical uncertainties. The systematic uncertainties, estimated separately for each asymmetry and hadron charge, are shown by the bands. The increase of Collins asymmetry at large $x$ and small $W$ increases with $Q^2$. For Sivers asymmetry, the non-negligible signal for positive hadrons at large $x$ is clearly visible, also at large $Q^2$, which is important for the Drell-Yan measurement foreseen by COMPASS in 2015. The other six asymmetries are compatible with zero.

\section{Conclusions}
\label{sec:conclusion}

A first study of the dependence of Collins, Sivers and the other six asymmetries upon kinematic variables $x$, $z$, $p_T$ and $W$ in different $Q^2$ regimes is performed, showing a non negligible $Q^2$ dependence of the Collins asymmetry. A more detailed multidimensional analysis of the eight asymmetries in simultaneous bins of ($x$,$Q^2$) and ($x$,$z$,$p_T$) is close to be finalised.

%
%



\begin{footnotesize}


%

\end{footnotesize}


\end{document}